\documentclass[%
 aip,
 jmp,%
 amsmath,amssymb,
 reprint,%
]{revtex4-1}

\usepackage{graphicx}
\usepackage{dcolumn}
\usepackage{bm}

\begin{document}

\preprint{AIP/123-QED}

\title[]{ON THE NOVEL MECHANISM OF ACCELERATION OF COSMIC PARTICLES.\\
Reports of Enlarged Session of the Seminar of I. Vekua Institute of
Applied Mathematics Volume 29, 2015}

\author{Z. Osmanov}
 \affiliation{School of Physics, Free University of Tbilisi.}
\email{z.osmanov@freeuni.edu.ge}



\date{\today}

\begin{abstract}
A novel model of particle acceleration in the rotating
magnetospheres of active galactic nuclei (AGN) and pulsars is
constructed. The particle energies may be boosted up to enormous
energies in a several step mechanism. In the first stage, the
Langmuir waves are centrifugally excited and amplified by means of a
parametric process that efficiently pumps rotational energy to
excite electrostatic fields. By considering the pulsars it is shown
that the Langmuir waves very soon Landau damp on the relativistic
electrons already present in a magnetosphere. It has been found that
the process is so efficient that no energy losses might affect the
mechanism of particle acceleration. Applying typical parameters for
young pulsars we have shown that by means of this process the
electrons might achieve energies of the order of $10^{18}$ eV. The
situation in AGN magnetospheres is slightly different. In the second
stage, the process of "Langmuir collapse" develops, creating
appropriate conditions for transferring electric energy to boost up
already high proton energies to much higher values. As in the
previous case, one can show that various energy losses are
relatively weak, and do not impose any significant constraints on
maximum achievable proton energies of the order of $10^{21}$ eV.
\end{abstract}

\maketitle

\section{Introduction}

It is observationally evident that energies of cosmic rays (CR)
range from several GeV, to thousands of EeV, therefore the major
question concerns the origin of such enormous energies. Generally
speaking it is strongly believed that Gamma ray bursts (GRB), AGN
and pulsars might be responsible for high and ultra-high energy
(UHE) CRs. In this context one of the puzzling and enigmatic
problems, is a concrete mechanism that provides efficient
acceleration. In the last century Fermi has proposed a mechanism
(Fermi acceleration) [1], that in most of the astrophysical
scenarios works very well. Although, acceleration of leptons might
encounter serious problems concerning pre-acceleration [2-3] and in
certain cases it is necessary to find alternative ways of
acceleration.

It is worth noting that AGN and pulsars have rapidly rotating
magnetospheres that potentially might significantly influence
dynamics of charged particles. For typical AGN with mass $M_8\equiv
M/(10^8M_{\odot})$ ($M_{\odot}\approx 2\times 10^{33}g$ is the solar
mass) the equipartition magnetic field is of the order of
\begin{equation}
\label{mag} B\approx\sqrt{\frac{2L}{r^2c}}\approx
27.5\times\left(\frac{L}{10^{42}erg/s}\right)^{1/2}\times\frac{R_{lc}}{r}G,
\end{equation}
where $L$ is the bolometric luminosity of AGN, $R_{lc}=c/\Omega$ is
the light cylinder radius, $r$ is the distance from the black hole
and we have taken into account that the angular velocity of rotation
of the AGN magnetosphere is given by
\begin{equation}
\label{rotat} \Omega\approx\frac{a c^3}{GM}\approx
10^{-3}\frac{a}{M_8}rad/s^2.
\end{equation}
Generally speaking, magnetic field is thought to be huge if its
energy density exceeds that of plasmas
\begin{equation}
\label{corot} \frac{B^2}{8\pi}> n\epsilon.
\end{equation}
where $n$ is the particle number density and $\epsilon$ is energy of
a single particle. From the above expression it is clear that in AGN
magnetospheres corotation of particles takes place up to
$\epsilon\approx 700$GeV.

Similarly, by taking into account the magnetic field of pulsars
close to the light cylinder area
\begin{equation}
\label{mag1} B\approx 3.2\times
10^6\times\left(\frac{P}{1s}\right)^{1/2}\times\left(\frac{\dot{P}}{10^{-15}ss^{-1}}
\right)^{1/2}G,
\end{equation}
and the Goldreich-Julian number density of electrons,
$n_{_{GJ}}=B/Pce$, where $P$ is the pulsar period, $c$ is the speed
of light and $e$ is the electron's charge, one can straightforwardly
show that condition of corotation (see Eq. (\ref{corot})) is
satisfied for a broad range of energies.

\section{Direct centrifugal acceleration}

We see that corotation is maintained up to the light cylinder
surface in the magnetospheres of AGN and pulsars. Therefore, it is
interesting to estimate efficiency of a centrifugal mechanism
developed in [4]. In particular, in [2] and [5] we have studied
efficiency of centrifugal acceleration in the magnetospheres of AGN
and pulsars respectively. Dynamics has been studied for electrons
"sliding" along corotating straight magnetic field lines inclined
with respect to the rotation axis. It has been shown that the
Lorentz factors behave as to be
\begin{equation}
\label{lorentz1} \gamma = \frac{\gamma_0}{1-\frac{r^2}{R_{lc}^2}},
\end{equation}
where $\gamma_0$ is the initial Lorentz factor of the electron. It
is clear from the above equation that on the light cylinder it
behaves asymptotically, indicating the existence of certain limiting
factors. In general one can show that acceleration lasts until the
particle encounters a soft photon, limiting the maximum attainable
energy (Inverse Compton (IC) mechanism) [2]
\begin{equation}
\label{gic} \gamma_{_{IC}}\approx\left(\frac{6\pi m_ec^4}{\sigma_T
L\Omega}\right)^2,
\end{equation}
where $m_e\approx 9,1\times 10^{-27}$g is electron's mass, $L$ is
the bolometric luminosity of a source and $\sigma_T\approx
6.65\times 10^{-25}$cm$^{-2}$ is the Thomson cross-section. Another
mechanism that potentially might limit the maximum energies of
electrons is the so called Breakdown of the bead on the wire (BBW)
approximation, leading to the following upper limit [2]
\begin{equation}
\label{gfb}
\gamma_{_{BBW}}\approx\frac{1}{c}\left(\frac{e^2L}{2m_e}\right)^{1/3}.
\end{equation}

It is clear that the real maximum attainable Lorentz factor should
satisfy the condition $\gamma=min \{\gamma_{_{IC}},
\gamma_{_{BBW}}\}$, which For AGN  leads to the maximum energy of
the order of $20$TeV [2] and in case of Crab-like pulsars the value
is even less $1$TeV [5].

As we see direct centrifugal acceleration has limits and cannot
explain ultra-high energies of cosmic rays.

\section{Acceleration via damping of centrifugally
excited Langmuir waves}

In this section we consider a slightly moderate mechanism of
acceleration. In particular, in [6,7] we have shown that rotation
energy might be very efficiently pumped into that of Langmuir waves.
On the next stage, under certain conditions, energy of electrostatic
modes might be converted into energy of plasma particles.

In this approach magnetospheric plasmas are considered to be
composed of two species: relatively low energy particles and high
energy particles. The process is governed by the system of
equations: Euler equation, continuity equation and Poisson equation,
which after Fourier transforming in time lead to the "dispersion
relation" of the centrifugally excited electrostatic modes [8]
\begin{equation}
\label{disp} \omega^2 -\omega_1^2 - \omega_2^2  J_0^2(b)= \omega_2^2
\sum_{\mu} J_{\mu}^{2}(b) \frac{\omega^2}{(\omega-\mu\Omega)^2},
\end{equation}
where $\omega_{1,2}\equiv\sqrt{4\eta\pi
e^2n_{1,2}/m_{1,2}\gamma_{1,2}^3}$ are the relativistic plasma
frequencies, $\gamma_{1,2}$ are the Lorentz factors, $m_{1,2}$ -
masses for the corresponding components respectively, $b=
{2ck}\sin\phi$, $\phi$ is half of the phase difference between
different species, $J_{\mu}(x)$ is the Bessel function of the first
kind and $\eta = 1$ for electron-proton plasmas and $\eta = 2$ for
electron-positron plasmas. The growth rate of the instability is
given by
\begin{equation}
 \label{grow}
 \Gamma= \frac{\sqrt3}{2}\left (\frac{\omega_1 {\omega_2}^2}{2}\right)^{\frac{1}{3}}
 {J_{\mu_{res}}(b)}^{\frac{2}{3}},
\end{equation}
where $\mu_{res}\equiv\omega_1/\Omega$. For typical values of
magnetospheric parameters of AGN: $\gamma_1\sim 10^6$, $\gamma_2\sim
10^3$ one can straightforwardly show that $1/\Gamma$ is less than
the kinematic timescale, $2\pi/\Omega$. The similar situation is
found in the case of pulsars, indicating that the energy pumping
from rotation into Langmuir waves is extremely efficient.

For these waves to effectively transfer their energy to particles,
the waves phase velocity must be close to the speed of light.
Further, in the vicinity of $\upsilon_{ph}$, there should be more
particles a little slower than the wave than particles which are a
little faster . For the given problem it is always possible to
situate $\upsilon_{ph}$ in the desired part of the primary beam
spectrum. Since the distribution function decreases with the Lorentz
factor, the number of electrons with $\upsilon_b<\upsilon_{ph}$
exceeds that of the electrons with $\upsilon_b>\upsilon_{ph}$, where
$\upsilon_b$ denotes the electron speed. Thus the optimum conditions
for effective Landau damping will pertain.

It has been shown that the total energy gained by the beam electrons
in magnetospheres of pulsars is given by [9]
\begin{equation}
\label{en}  \epsilon\approx \frac{n_1F\delta r}{n_{_{GJ}}},
\end{equation}
where $\delta r\sim c/\Gamma$, $F\approx 2m_ec\Omega\xi (r)^{-3}$,
$\xi (r) = \left(1-\Omega^2r^2/c^2\right)^{1/2}$. It has been shown
that the Crab-like pulsars might provide with energies of the order
$400$TeV, whereas in the magnetospheres of newly born millisecond
pulsars electrons might be accelerated up to $10^{18}$eV.

Since the AGN magnetospheres are predominantly composed of protons
and electrons, it has been found that the efficient Langmuir
collapse develops after centrifugally inducing electrostatic waves.
This process leads to even higher accumulation of energy, finally
transformed to energy of protons via the Landau damping. The
corresponding expression of proton energy is given by
\begin{equation}
\label{energy1} \epsilon_p\left(eV\right)\approx 2\times
10^{20}\times\left(\frac{10^2}{\gamma_2}\right)^5 \times
M_8^{-5/2}\times L_{43}^{5/2},
\end{equation}
where $L_{43}\equiv L/10^{43}$erg/s is the dimensioneless luminosity
of AGN. As it is clear the present mechanism can explain energies of
cosmic rays. Therefore, it would be interesting to further develop
our model to apply for neutrino astrophysics.

\section*{Acknowledgments}
The research was partially supported by the Shota Rustaveli National
Science Foundation grant (N31/49).



\thebibliography{aipsamp}
\bibitem{1} Fermi E. On the Origin of the Cosmic Radiation. {\it Phys. Rev.}
\textbf{75}, (2000), 1169-1174.

\bibitem{2} Osmanov Z., Rogava A. \& Bodo G. On the efficiency of particle
acceleration by rotating magnetospheres in AGN. {\it A\&A},
\textbf{470}, (2007) 395-400.

\bibitem{3} Catanese M. \& Weeks T. C. Constraints on Cosmic-Ray Origin
Theories From TeV Gamma-Ray Observations. {\it PASP}, \textbf{111},
(1999) 170-177.

\bibitem{4} Machabeli G. Z. \& Rogava, A. D. Centrifugal force: A gedanken
experiment. {\it Phys. Rev. A}, \textbf{50}, (1994) 98-103

\bibitem{5} Osmanov Z. \& Rieger F. On particle acceleration and very high
energy gamma-ray emission in Crab-like pulsars. {\it A\&A},
\textbf{502}, (2009) 15-20.

\bibitem{6} Machabeli G. Z., Osmanov Z. N. \& Mahajan S. Parametric mechanism
of the rotation energy pumping by a relativistic plasma. {\it PhPl},
\textbf{12}, (2005) 1-6.

\bibitem{7} Osmanov Z. Centrifugally driven electrostatic instability in
extragalactic jets. {\it PhPl}, \textbf{15}, (2008) 1-7

\bibitem{8} Osmanov Z., Mahajan S., Machabeli G. \& Chkheidze N. Extremely
efficient Zevatron in rotating AGN magnetospheres. {\it MNRAS},
\textbf{445}, (2014) 4155-4160

\bibitem{9} Mahajan S., Machabeli G., Osmanov Z. \& Chkheidze N. Ultra High
Energy Electrons Powered by Pulsar Rotation. {\it NatSR},
\textbf{3}, (2014) 1-5
\end{document}